\documentstyle[amsmath,amssymb,12pt]{article}
\newtheorem{th}{Theorem}[section] \newtheorem{lm}[th]{Lemma}
 
\newcommand{\ket}[1]{\left | #1 \right \rangle}

 \newcommand{\norm}[1]{\left\lVert #1 \right\rVert}
\newcommand{\rr}{\mbox{$\mathbb R$}} \newcommand{\cc}{\mbox{$\mathbb
C$}} \newcommand{\ov}{\overline} 
 \newcommand{\ft}{\footnotesize}

\begin{document} 

\title{\bf An Almost-Quadratic Lower Bound for Quantum Formula Size
\thanks{This work was supported in part by grants from the Revolutionary Computing
group at JPL (contract \#961360), and from the DARPA Ultra program
(subcontract from Purdue University \#530--1415--01).}} 

\author{Vwani P. Roychowdhury\thanks{e--mail: {\tt vwani@ee.ucla.edu}} 
\hspace{2cm} Farrokh Vatan\thanks{e--mail: {\tt vatan@ee.ucla.edu}} \\
Electrical Engineering Department\\   UCLA\\   Los Angeles, CA 90095} 

\date{ } 

\maketitle 

\begin{abstract} We show that Nechiporuk's method \cite{wegener} for proving
lower bound for Boolean formulas can be extended to the quantum case.
This leads to an $\Omega(n^2/\log^2 n)$ lower bound for quantum formulas
computing an explicit function. The only known previous explicit 
lower bound for quantum formulas \cite{yao} states that the majority function 
does not have a linear--size quantum formula. \end{abstract} 

\section{Introduction} 

Computational devices based on quantum physics have attracted much
attention lately, and quantum algorithms that perform much faster than
their classical counterparts have been developed
\cite{grover,shor,simon}. To provide a systematic study of the
computational power of quantum devices, models similar to those for
classical computational devices have been proposed. Deutsch
\cite{deutsch85} formulated the notion of quantum Turing machine. This
approach was further developed by Bernstein and Vazirani \cite{vazirani},
and the concept of an efficient universal quantum Turing machine was
introduced. As in the case of classical Boolean computation, there is
also a quantum model of computation based on circuits (or networks). Yao
\cite{yao} proved that the quantum circuit model, first introduced by
Deutsch \cite{deutsch89}, is equivalent to the quantum Turing machine model. 

Since every Boolean circuit can be simulated by a quantum circuit, with
at most a polynomial factor increase in its size, any nontrivial lower bound 
for quantum circuits could have far reaching consequences. In classical
Boolean circuit theory, all nontrivial lower bounds are for proper
subclasses of Boolean circuits such as monotone circuits, formulas,
bounded-depth circuits, etc. In the quantum case also it seems that 
the only hope to prove nontrivial lower bounds is for proper subclasses of
quantum circuits. So far the only such known lower bound has been derived by Yao
\cite{yao} for quantum formulas\footnote{There are exponential lower
bounds on the time of quantum computation for the black--box model 
(see, e.g., \cite{beals}), but
they do not apply to the size of quantum circuits.}. The quantum formula is a
straightforward generalization of the classical Boolean formula: in both
cases, the graph of the circuit is a tree. Yao has proved that the
quantum formula size of the majority function $\mbox{MAJ}_n$ is not
linear\footnote{The value of $\mbox{MAJ}_n(x_1,\ldots,x_n)$ is $1$ if at
least $\lceil n/2\rceil$ of inputs are 1.}; i.e., if $L(\mbox{MAJ}_n)$ 
denotes the minimum quantum formula size of $\mbox{MAJ}_n$ then
$\lim_{n\longrightarrow\infty}L(\mbox{MAJ}_n)/n=\infty$. This bound is
derived from a bound on the quantum communication complexity of Boolean
functions. 

In this paper, we prove an almost quadratic lower bound for quantum
formula size. The key step in the derivation of this lower bound is the
extension of Nechiporuk's method to quantum formulas; for a detailed
discussion of Nechiporuk's method in the Boolean setting see
\cite{dunne,wegener}. Nechiporuk's method has been used in several
different areas of Boolean complexity (e.g., see \cite{dunne} for details). It
has also been applied to models where the gates do not take on binary or
discrete values, but the input/output map still corresponds to a Boolean
function. For example, in \cite{turan} this method has been used to get
a lower bound for arithmetic and threshold formulas. The challenging
part of this method is a step that we shall refer to as ``path
squeezing'' (see Section \ref{lowerbound} for the exact meaning of it). 
Although in the
case of Boolean gates, this part can be solved easily, in the case of
analog circuits it is far from obvious (see \cite{turan}). For the
quantum formulas ``path squeezing'' becomes even more complicated,
because here we should take care of any {\em quantum entanglement} and
interference phenomenon. We show that it is still possible to squeeze a
path with arbitrary number of constant inputs to a path
with a fixed number of inputs. This leads to a lower bound of $\Omega(n^2/\log^2 n)$ 
on the size of quantum formulas computing a class of explicit functions. 
For example, we get such a bound for the Element Distinctness function $\mbox{ED}_n$.
The input of $\mbox{ED}_n$ is of the form $(z_1,\ldots,z_\ell)$, where each $z_j$ is
a string of $2\log\ell$ bits. Then $\mbox{ED}_n(z_1,\ldots,z_\ell)=1$ if and only
if all these strings are pairwise distinct.

In this paper we use the notation $|\cdot|$ for two different purposes. When $\alpha$ 
is a complex number, $|\alpha|$ denotes the absolute value of $\alpha$; i.e.,
$|\alpha|=\sqrt{\alpha\cdot\alpha ^*}$. While if $X$ is a set then $|X|$
denotes the cardinality of $X$.

\section{Preliminaries} 

A {\em quantum circuit} is defined as straightforward generalization of
acyclic classical (Boolean) circuit (see \cite{deutsch89}). 
For constructing a quantum circuit, we begin with a {\em basis} of
quantum gates as elementary gates.
Each such elementary gate with $d$ inputs and outputs
is uniquely represented by a unitary operation on $\cc^{2^d}$. The gates 
are interconnected by quantum ``wires''. Each wire
represents a quantum bit, {\em qubit}, which is a 2--state quantum
system represented by a unit vector in $\cc^2$. Let
$\{\ket{0},\ket{1}\}$ be the standard orthonormal basis of $\cc^2$. The
$\ket{0}$ and $\ket{1}$ values of a qubit correspond to the classical
Boolean $0$ and $1$ values, but a qubit can also be in a superposition
of the form $\alpha\ket{0}+\beta\ket{1}$, where $\alpha,\beta\in\cc$ and
$|\alpha|^2+|\beta|^2=1$. Each gate $g$ with $d$ inputs represents a unitary
operation $U_g\in\mbox{U}(2^d)$. Note that the output of such gate, in general,
is not a tensor product of its inputs, but an {\em entangled state}; e.g., a
state like $\frac{1}{\sqrt{2}}\ket{00}+\frac{1}{\sqrt{2}}\ket{11}$ which can not
be written as a tensor product.

If the circuit has $m$ inputs, then for each $d$--input gate $g$, the unitary
operation $U_g\in\mbox{U}(2^d)$ can be considered in a natural way as an
operator in $\mbox{U}(2^m)$ by acting as the identity operator on the
other $m-d$ qubits. Hence, a quantum circuit with $m$ inputs computes a
unitary operator in $\mbox{U}(2^m)$, which is the product of successive
unitary operators defined by successive gates. 

In this paper, we consider quantum circuits that compute Boolean
functions. Consider a quantum circuit $C$ with $m$ inputs. Suppose that
$C$ computes the unitary operator $U_C\in\mbox{U}(2^m)$. We say $C$
computes the Boolean function $f\colon\{0,1\}^n\longrightarrow \{0,1\}$
if the following holds. The inputs are labeled by the variables
$x_1,x_2,\ldots, x_n$ or the constants $\ket{0}$ or $\ket{1}$ (different
inputs may be labeled by the same variable $x_j$). We consider one of
the output wires, say the first one, as the output of the
circuit. To compute the value of the circuit at
$\alpha=(\alpha_1,\ldots,\alpha_n)\in\{0,1\}^n$, let the value of each
input wire with label $x_j$ be $\ket{\alpha_j}$. These inputs, along
with the constant inputs to the circuit, define a vector $\ket{\alpha}$ 
in $\cc^{2^m}$. In fact this
vector is a standard basis vector of the following form (up to some
repetitions and a permutation) \[
\ket{\alpha}=\ket{\alpha_1}\otimes\cdots\otimes\ket{\alpha_n}\otimes
\ket{0}\otimes\cdots\otimes\ket{1} \] The act of the circuit $C$ on the
input $\ket{\alpha}$ is the same as $U_C(\ket{\alpha})$. Note that since
$U_C$ is unitary, $\norm{U_C(\ket{\alpha})}=1$. We decompose the vector
$U_C(\ket{\alpha})\in\cc^{2^m}$ with respect to the output qubit. Let
the result be \[ U_C(\ket{\alpha}) = \ket{0}\otimes
\ket{A_0}+\ket{1}\otimes \ket{A_1} .\] Then we define the {\em
probability} that $C$ outputs 1 (on the input $\alpha$) as
$p_\alpha=\norm{\ket{A_1}}^2$, i.e., the square of the length of
$\ket{A_1}\in\cc^{2^{m-1}}$. Finally, we say that the quantum circuit
$C$ computes the Boolean function $f$ if for every $\alpha\in\{0,1\}^n$,
if $f(\alpha)=1$ then $p_\alpha > 2/3$ and if $f(\alpha)=0$ then
$p_\alpha < 1/3$. 

\begin{figure} \begin{center} \unitlength=.5mm \begin{tabular}{cp{8mm}c}
\begin{picture}(110,50)(0,-5) \put(0,0){\makebox(0,0){$x_1$}}
\put(0,15){\makebox(0,0){$x_2$}} \put(0,30){\makebox(0,0){$0$}}
\put(0,45){\makebox(0,0){$x_1$}} \put(5,15){\vector(1,0){10}}
\put(5,30){\vector(1,0){10}} \put(5,45){\vector(1,0){10}}
\put(19,30){\makebox(0,0){\framebox(8,34){$\ft g_1$}}}
\put(5,0){\vector(1,0){28}} \put(23,15){\vector(1,0){10}}
\put(37,7.5){\makebox(0,0){\framebox(8,19){$\ft g_2$}}}
\put(41,15){\vector(1,0){10}} \put(23,30){\vector(1,0){28}}
\put(55,22.5){\makebox(0,0){\framebox(8,19){$\ft g_3$}}}
\put(41,0){\vector(1,0){46}} \put(59,15){\vector(1,0){28}}
\put(59,30){\vector(1,0){10}} \put(23,45){\vector(1,0){46}}
\put(73,37.5){\makebox(0,0){\framebox(8,19){$g_4$}}}
\put(77,30){\vector(1,0){10}} \put(77,45){\vector(1,0){10}}
\put(99,45){\makebox(0,0){\ft output}} \end{picture} 

& & 

\begin{picture}(100,50)(0,-5) \put(0,0){\makebox(0,0){$x_1$}}
\put(0,15){\makebox(0,0){$x_2$}} \put(0,30){\makebox(0,0){$0$}}
\put(0,45){\makebox(0,0){$x_1$}} \put(7,0){\circle*{4}}
\put(7,15){\circle*{4}} \put(7,30){\circle*{4}} \put(7,45){\circle*{4}}
\put(7,45){\line(1,0){90}} \put(37,45){\circle*{4}}
\put(37,51){\makebox(0,0){$\ft g_1$}} \put(97,45){\circle*{4}}
\put(97,51){\makebox(0,0){$\ft g_4$}} \put(7,30){\line(2,1){30}}
\put(7,15){\line(1,1){30}} \put(7,0){\line(1,0){60}}
\put(67,0){\circle*{4}} \put(67,-6){\makebox(0,0){$\ft g_2$}}
\put(67,0){\line(2,3){30}} \put(67,0){\line(-2,3){30}}
\put(82,22.5){\circle*{4}} \put(89,22.5){\makebox(0,0){$\ft g_3$}}
\put(82,22.5){\line(-2,1){45}} 

\end{picture} 

\\ \vspace{3mm} & \\ 

\begin{picture}(110,50)(0,-5) \put(0,0){\makebox(0,0){$x_1$}}
\put(0,15){\makebox(0,0){$x_2$}} \put(0,30){\makebox(0,0){$0$}}
\put(0,45){\makebox(0,0){$x_1$}} \put(5,15){\vector(1,0){10}}
\put(5,30){\vector(1,0){10}} \put(5,45){\vector(1,0){10}}
\put(19,30){\makebox(0,0){\framebox(8,34){$\ft g_1$}}}
\put(5,0){\vector(1,0){28}} \put(37,0){\makebox(0,0){\framebox(8,9){$\ft
g_2$}}} \put(41,0){\vector(1,0){10}} \put(23,15){\vector(1,0){28}}
\put(55,7.5){\makebox(0,0){\framebox(8,19){$\ft g_3$}}}
\put(59,0){\vector(1,0){28}} \put(59,15){\vector(1,0){28}}
\put(23,30){\vector(1,0){46}} \put(23,45){\vector(1,0){46}}
\put(73,37.5){\makebox(0,0){\framebox(8,19){$g_4$}}}
\put(77,30){\vector(1,0){10}} \put(77,45){\vector(1,0){10}}
\put(99,15){\makebox(0,0){\ft output}} \end{picture} 

& & 

\begin{picture}(100,50)(0,-5) \put(0,0){\makebox(0,0){$x_1$}}
\put(0,15){\makebox(0,0){$x_2$}} \put(0,30){\makebox(0,0){$0$}}
\put(0,45){\makebox(0,0){$x_1$}} \put(7,0){\circle*{4}}
\put(7,15){\circle*{4}} \put(7,30){\circle*{4}} \put(7,45){\circle*{4}}
\put(7,45){\line(1,0){90}} \put(37,45){\circle*{4}}
\put(37,51){\makebox(0,0){$\ft g_1$}} \put(97,45){\circle*{4}}
\put(97,51){\makebox(0,0){$\ft g_3$}} \put(7,30){\line(2,1){30}}
\put(7,15){\line(1,1){30}} \put(7,0){\line(1,0){30}}
\put(37,0){\circle*{4}} \put(37,-6){\makebox(0,0){$\ft g_2$}}
\put(37,0){\line(4,3){60}} 

\end{picture}\end{tabular}\end{center}
\caption{Quantum circuits and their computation graphs; the top circuit
is not a formula while the bottom one is a formula.} 
\label{fig1} \end{figure}
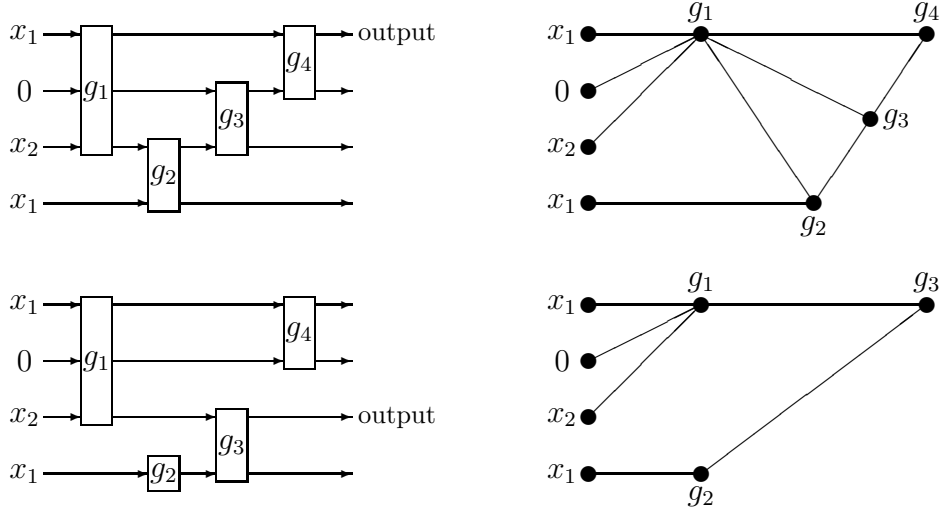

Following Yao \cite{yao}, we define quantum formulas as a subclass of
quantum circuits. A quantum circuit $C$ is a {\em formula} if for every
input there is a unique path that connects it to the output qubit. To
make this definition more clear we define the {\em computation graph} of
$C$, denoted by $G_C$. The nodes of $G_C$ correspond to a subset of the gates of
$C$. We start with the output gate of $C$, i.e., the gate which provides
the output qubit, and let it be a node of $G_C$. Once a node $v$ belongs
to $G_C$ then all gates in $C$ that provide inputs to $v$ are considered
as adjacent nodes of $v$ in $G_C$. Then $C$ is a formula if the graph
$G_C$ is a tree. Figure \ref{fig1} provides examples of quantum circuits of both
kinds, i.e., circuits that are also quantum formulas, and circuits that
are not formulas. 

All circuits that we consider are over some fixed universal quantum
basis. The lower bound does not depend on the basis; the only condition
is that the number of inputs (and so the number of outputs) of each gate
be bounded by some fixed constant number (this condition is usually
considered as part of the definition of a quantum basis). For example,
this basis can be the set of all 2--input 2--output quantum gates, and as
as it is shown in \cite{barenco-etal}, this basis is universal. 

For our proof we also need a Shannon--type result for quantum circuits.
Knill \cite{knill} has proved several theorems about the quantum circuit
complexity of almost all Boolean functions. We will use the following
theorem. 

\begin{th} 
{\em \cite{knill}} The number of different $n$--variable
Boolean functions that can be computed by size $N$ quantum circuits ($n\leq
N$) with $d$--input $d$--output elementary gates is at most $2^{cN\log
N}$, where $c$ is a constant which is a function of $d$. 
\label{knill} 
\end{th} 

For the sake of completeness, in the Appendix we have provided a proof for a
slightly weaker bound. Our approach is different from that in \cite{knill} and it seems it
is shorter and simpler than his proof. Although the bound that we get is a little
weaker than the bound provided by the above theorem (it is of the form $2^{O(nN)}$),
our bound results in the same bound if $\log(N)=\Omega(n)$. Thus our result provides
the same bound for the complexity of the most 
difficult function and the bound we get in this paper.

We also need to consider orthonormal bases in the space $\cc^{2^n}$
other than the standard basis. In the context of quantum physics, we
identify the Hilbert space $\cc^{2^n}$ as the tensor product space
$\bigotimes_{j=1}^n \cc^2$, and the standard basis consists of the
vectors \[
\ket{c_1}\otimes\ket{c_2}\otimes\cdots\otimes\ket{c_n}=\ket{c_1c_2\cdots
c_n} , \qquad c_j\in\{0,1\} . \] The next lemma provides a method to
construct other sets of mutually orthogonal unit vectors in $\cc^{2^n}$.

\begin{lm} Let $\ket{A_j}\in\cc^{2^k}$ and $\ket{B_j}\in\cc^{2^m}$ for
$j=1,2$. Then for the length and inner product of the vectors
$\ket{A_j}\otimes\ket{B_{j}}\in\cc^{2^{k+m}}$ we have 
\[\bigl\lVert\ket{A_{1}}\otimes
\ket{B_{1}}\bigr\rVert=\bigl\lVert\ket{A_{1}}\bigr\rVert
\cdot\bigl\lVert\ket{B_{1}}\bigr\rVert , \] 
and if $\ket{x_1}=\ket{A_1}\otimes\ket{B_1}$ and 
$\ket{x_2}=\ket{A_2}\otimes\ket{B_2}$ then
\[\bigl\langle x_1\bigl\lvert x_2\bigr\rangle\bigr . = \bigl\langle A_1
\bigl\lvert A_2 \bigr\rangle\bigr . \cdot \bigl\langle B_1 \bigl\lvert
B_2 \bigr\rangle\bigr . . \] 
\label{orthonormal} \end{lm} 

{\bf Proof.} Suppose that
$\ket{A_{1}}=\sum_{c\in\{0,1\}^k}\alpha_c\ket{c}$ and
$\ket{B_{1}}=\sum_{d\in\{0,1\}^m}\beta_d\ket{d}$. Then \begin{eqnarray*}
\bigl\lVert\ket{A_{1}}\otimes\ket{B_{1}}\bigr\rVert & = &
\biggl\lVert\sum_{c\in\{0,1\}^k}\sum_{d\in\{0,
1\}^m}\alpha_c\beta_d\ket{c} \otimes\ket{d}\biggr\rVert \\ & = &
\sum_{c\in\{0,1\}^k}\sum_{d\in\{0,1\}^m}|\alpha_c|^2 |\beta_d|^2 \\ & =
& \biggl ( \sum_{c\in\{0,1\}^k}|\alpha_c|^2 \biggr ) \cdot \biggl (
\sum_{d\in\{0,1\}^m}|\beta_d|^2 \biggr ) \\ & = &
\bigl\lVert\ket{A_{1}}\bigr\rVert \cdot\bigl\lVert\ket{B_{1}}\bigr\rVert
. \end{eqnarray*} The proof in the case of the inner product is similar.
$\Box$ 

\vspace{6mm} The above lemma can easily be generalized to the families 
of more than two vectors, and the generalized version is stated below. 

\begin{lm} Let $\ket{A_j}\in\cc^{2^k}$ and $\ket{B_\ell}\in\cc^{2^m}$ be
unit vectors (for $j$ and $\ell$ in some index sets). If $\ket{A_j}$ are
pairwise orthogonal and $\ket{B_\ell}$ are pairwise orthogonal then the
family \[ \left\{ \ket{A_j}\otimes\ket{B_{\ell}}\in\cc^{2^{k+m}} \colon
j,\ell\right\} \] is an orthonormal set. \label{basis} \end{lm} 

The following lemma, although seemingly obvious, is crucial for the ``path squeezing'' 
technique in the proof of the lower bound.  

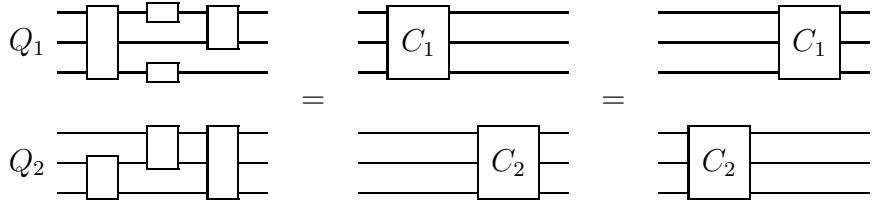
\begin{figure} \begin{center} \unitlength=.4mm 

\begin{picture}(270,60)(0,0) 
\put(-10,10){\makebox(0,0){$Q_2$}}
\put(-10,50){\makebox(0,0){$Q_1$}} \put(0,0){\line(1,0){10}}
\put(0,10){\line(1,0){10}} \put(0,20){\line(1,0){30}}
\put(0,40){\line(1,0){10}} \put(0,50){\line(1,0){10}}
\put(0,60){\line(1,0){10}} \put(15,50){\makebox(0,0){\framebox(10,24){
}}} \put(15,5){\makebox(0,0){\framebox(10,14){ }}}
\put(20,0){\line(1,0){30}} \put(20,10){\line(1,0){10}}
\put(20,40){\line(1,0){10}} \put(20,50){\line(1,0){30}}
\put(20,60){\line(1,0){10}} \put(35,15){\makebox(0,0){\framebox(10,14){
}}} \put(35,40){\makebox(0,0){\framebox(10,6){ }}}
\put(35,60){\makebox(0,0){\framebox(10,6){ }}}
\put(40,10){\line(1,0){10}} \put(40,20){\line(1,0){10}}
\put(40,40){\line(1,0){30}} \put(40,60){\line(1,0){10}}
\put(55,10){\makebox(0,0){\framebox(10,24){ }}}
\put(55,55){\makebox(0,0){\framebox(10,14){ }}}
\put(60,0){\line(1,0){10}} \put(60,10){\line(1,0){10}}
\put(60,20){\line(1,0){10}} \put(60,50){\line(1,0){10}}
\put(60,60){\line(1,0){10}} 

\put(85,30){\makebox(0,0){$=$}} 

\put(100,0){\line(1,0){40}} \put(100,10){\line(1,0){40}}
\put(100,20){\line(1,0){40}} \put(100,40){\line(1,0){10}}
\put(100,50){\line(1,0){10}} \put(100,60){\line(1,0){10}}
\put(120,50){\makebox(0,0){\framebox(20,24){$C_1$}}}
\put(150,10){\makebox(0,0){\framebox(20,24){$C_2$}}}
\put(160,0){\line(1,0){10}} \put(160,10){\line(1,0){10}}
\put(160,20){\line(1,0){10}} \put(130,40){\line(1,0){40}}
\put(130,50){\line(1,0){40}} \put(130,60){\line(1,0){40}} 

\put(185,30){\makebox(0,0){$=$}} 

\put(200,0){\line(1,0){10}} \put(200,10){\line(1,0){10}}
\put(200,20){\line(1,0){10}} \put(200,40){\line(1,0){40}}
\put(200,50){\line(1,0){40}} \put(200,60){\line(1,0){40}}
\put(250,50){\makebox(0,0){\framebox(20,24){$C_1$}}}
\put(220,10){\makebox(0,0){\framebox(20,24){$C_2$}}}
\put(230,0){\line(1,0){40}} \put(230,10){\line(1,0){40}}
\put(230,20){\line(1,0){40}} \put(260,40){\line(1,0){10}}
\put(260,50){\line(1,0){10}} \put(260,60){\line(1,0){10}} 

\end{picture} \end{center} 
\caption{Decomposition of a quantum subcircuit acting on disjoint sets of
qubits (Lemma~\protect\ref{composition} (a)).} \label{fig2} \end{figure}

\begin{figure} \begin{center} \unitlength=.4mm 

\begin{picture}(300,110)(0,-80) 
\put(0,0){\vector(1,0){30}}\put(30,16){\line(-1,1){25}}
\put(40,8){\makebox(0,0){\framebox(20,24){$g_1$}}}
\put(10,5){\makebox(0,0){$q$}}
\put(30,16){\vector(1,-1){0}}\put(18,35){\makebox(0,0){$r_1$}}

\put(50,0){\vector(1,0){30}}\put(80,16){\line(-1,1){25}}
\put(90,8){\makebox(0,0){\framebox(20,24){$g_2$}}}
\put(70,5){\makebox(0,0){$q$}}
\put(80,16){\vector(1,-1){0}}\put(68,35){\makebox(0,0){$r_2$}}
\put(50,16){\line(1,-1){90}}\put(140,-74){\vector(1,0){110}}
\put(100,16){\line(1,-1){70}}\put(170,-54){\vector(1,0){80}}

\put(140,0){\vector(1,0){30}}\put(170,16){\line(-1,1){25}}
\put(180,8){\makebox(0,0){\framebox(20,24){$g_t$}}}
\put(160,5){\makebox(0,0){$q$}}
\put(170,16){\vector(1,-1){0}}\put(158,35){\makebox(0,0){$r_t$}}
\put(190,0){\vector(1,0){30}}\put(210,5){\makebox(0,0){$q$}}
\put(190,16){\line(1,-1){40}}\put(230,-24){\vector(1,0){20}}
\put(270,-49){\makebox(0,0){\framebox(40,60)
{\scriptsize\begin{tabular}{c} postponed \\ gates\end{tabular}}}}

\put(120,8){\circle*{3}}\put(127,8){\circle*{3}}\put(134,8){\circle*{3}}
\put(240,-30){\circle*{3}}\put(240,-39){\circle*{3}}\put(240,-48){\circle*{3}}
\put(240,-80){\makebox(0,0){$r_1$}}\put(240,-60){\makebox(0,0){$r_2$}}
\put(240,-20){\makebox(0,0){$r_t$}}

\end{picture} \end{center} 
\caption{Postponing the gates (Lemma~\protect\ref{composition} (b)).} 
\label{fig3} \end{figure}
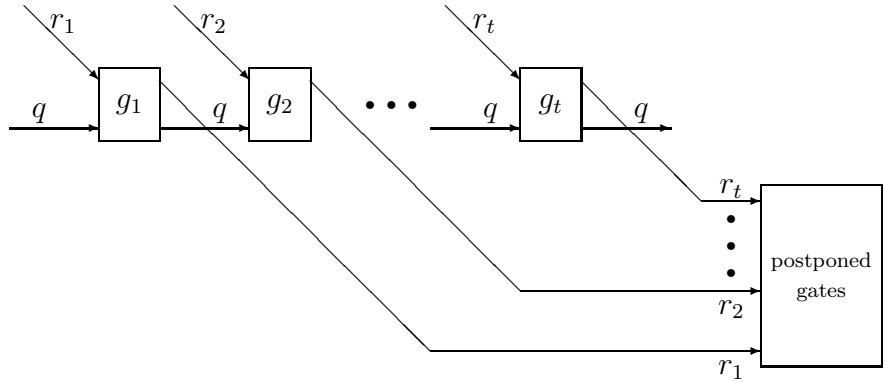

\begin{lm} (a) Suppose that $C$ is a subcircuit of a quantum circuit. Let the inputs of 
$C$ be divided into two disjoint sets of qubits $Q_1$ and $Q_2$. Suppose that each gate 
of $C$ either acts only on qubits from $Q_1$ or only on qubits from $Q_2$. Then there
are subcircuits $C_1$ and $C_2$ such that $C_j$ acts only on qubits from
$Q_j$ and the operation of $C$ is the composition of operations of $C_1$
and $C_2$ no matter in which order they act; i.e., $C=C_1\circ
C_2=C_2\circ C_1$. So the subcircuit $C$ can be substituted
by $C_1$ and $C_2$ (see Figure {\em \ref{fig2}}).

(b) Let $C$ be a quantum subcircuit with distinct input qubits $q$ and $r_1,\ldots,r_t$. 
Suppose that only $t$ gates $g_1,\ldots,g_t$ in $C$ act on $q$ and each $g_j$ acts 
on $q$ and $r_j$. Then, w.l.o.g., we can assume that each qubit $r_j$ after entering 
the gate $g_j$ will not interact with any other qubit until the gate $g_t$ is performed 
(see Figure {\em \ref{fig3}}).
\label{composition} \end{lm}

{\bf Proof.}
Part (a) is based on the following simple
observation. If $M\in\mbox{U}(2^m)$ and $N\in\mbox{U}(2^n)$ then
\[ M\otimes N=(M\otimes I_n)\circ(I_m\otimes N)=(I_m\otimes N)\circ(M\otimes I_n), \]
where $I_t$ is the identity map in $\mbox{U}(2^t)$.
Note that the inputs of the subcircuit $C$ may be in an entangled state; but
to see that the equality $C=C_1\circ C_2=C_2\circ C_1$ holds, it is enough to check this 
equality for the standard basis and extend it to the whole space by linearity.

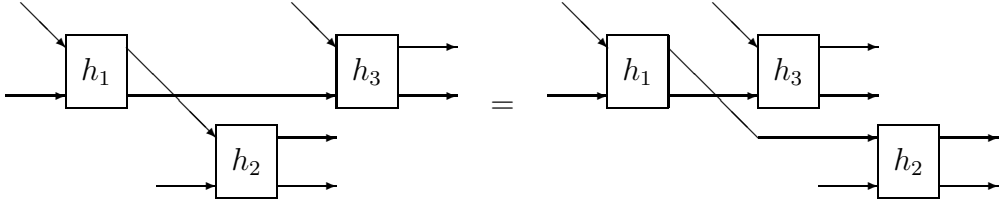
\begin{figure} \begin{center} \unitlength=.4mm 

\begin{picture}(330,50)(0,-30) 
\put(0,0){\vector(1,0){20}}\put(20,16){\line(-1,1){15}}
\put(30,8){\makebox(0,0){\framebox(20,24){$h_1$}}}
\put(20,16){\vector(1,-1){0}}
\put(40,16){\vector(1,-1){30}}\put(50,-30){\vector(1,0){20}}
\put(80,-22){\makebox(0,0){\framebox(20,24){$h_2$}}}
\put(40,0){\vector(1,0){70}}\put(110,16){\line(-1,1){15}}
\put(120,8){\makebox(0,0){\framebox(20,24){$h_3$}}}
\put(110,16){\vector(1,-1){0}}
\put(130,0){\vector(1,0){20}}\put(130,16){\vector(1,0){20}}
\put(90,-30){\vector(1,0){20}}\put(90,-14){\vector(1,0){20}}

\put(165,-4){\makebox(0,0){$=$}}

\put(180,0){\vector(1,0){20}}\put(200,16){\line(-1,1){15}}
\put(210,8){\makebox(0,0){\framebox(20,24){$h_1$}}}
\put(200,16){\vector(1,-1){0}} 
\put(220,0){\vector(1,0){30}}\put(250,16){\line(-1,1){15}}
\put(260,8){\makebox(0,0){\framebox(20,24){$h_3$}}}
\put(250,16){\vector(1,-1){0}}
\put(220,16){\line(1,-1){30}}\put(250,-14){\vector(1,0){40}}
\put(270,-30){\vector(1,0){20}}
\put(300,-22){\makebox(0,0){\framebox(20,24){$h_2$}}}
\put(270,0){\vector(1,0){20}}\put(270,16){\vector(1,0){20}}
\put(310,-30){\vector(1,0){20}}\put(310,-14){\vector(1,0){20}}

\end{picture} \end{center} 
\caption{Changing the order of gates (Lemma~\protect\ref{composition} (b)).} 
\label{fig4} \end{figure}

Part (b) follows simply from part (a); as in Figure \ref{fig4}, part (a) can be applied on
subcircuit consisting of gates $h_2$ and $h_3$. Note that in this case also input qubits $r_j$
of $g_j$'s may be in an entangled state. Again a linearity argument shows that we have to
consider only the case that $r_j$'s are in a product state. $\Box$

\section{The lower bound} 
\label{lowerbound}

Let $f(x_1,\ldots ,x_n)$ be a Boolean function, and let
$X=\{x_1,\ldots,x_n\}$ be the set of the input variables. Consider a partition
$\{S_1,\ldots,S_k\}$ of $X$; i.e., $X=\bigcup_{1\leq j\leq k}S_j$ and
$S_{j_1}\cap S_{j_2}=\emptyset$, for $j_1\neq j_2$. Let $n_j=|S_j|$, for
$j=1,\ldots,k$. Let $\Sigma_j$ be the set of all subfunctions of $f$ on
$S_j$ obtained by fixing the variables outside $S_j$ in all possible
ways. We denote the cardinality of $\Sigma_j$ by $\sigma_j$. 

As an example, we compute the above parameters for the Element Distinctness
function $\mbox{ED}_n$ (see \cite{boppana}). 
Let $n=2\ell\log\ell$ (so $\ell=\Omega(n/\log
n)$) and divide the $n$ inputs of the function into $\ell$ strings each
of $2\log\ell$ bits. Then the value of $\mbox{ED}_n$ is 1 if and only if these
$\ell$ strings are pairwise distinct. We consider the partition 
$(S_1,\ldots,S_\ell)$ such that each $S_j$ contains all variables of the
same string. Thus $n_j=|S_j|=2\log\ell$. Each string in $S_j$ represents an
integer from the set $\{0,1,\ldots,\ell^2-1\}$. The function $\mbox{ED}_n$ is
symmetric with respect to $S_j$'s; so $|\Sigma_j|=|\Sigma_{j'}|$. To estimate
$|\Sigma_1|$, note that if the strings $(z_2,\ldots,z_\ell)$
in $S_2,\ldots,S_\ell$ represent distinct integers then the corresponding 
subfunction is different from any subfunction corresponding to any other string.
So $\sigma_j=|\Sigma_1|\geq\binom{\ell^2}{\ell-1} > \ell^{\ell-1}$.

\begin{th} Every quantum formula computing $f$ has size \[ \Omega\biggl
( \sum_{1\leq j\leq k}\frac{\log(\sigma_j)}{\log\log(\sigma_j)}\biggr )
. \] \label{bound} \end{th} 

{\bf Proof.} We give a proof for any basis consisting of 2--input 2--output quantum 
gates. The proof for the other bases is a simple generalization of this proof.

Let $F$ be a formula computing $f$. Let $\ov{S_j}$ be the
set of input wires of $F$ labeled by a variable from $S_j$, and let
$s_j=|\ov{S_j}|$. Then 
\begin{equation} 
\mbox{\sf size}(F)=\Omega\biggl( \sum_{1\leq j\leq k}s_j\biggr ) . 
\label{size} \end{equation} 
We want
to consider the formulas obtained from $F$ by letting the variable
inputs not in $\ov{S_j}$ to some constant value $\ket{0}$ or $\ket{1}$.
In this regard, let $P_j$ be the set of all paths from an input wire in
$\ov{S_j}$ to the output of $F$. Finally, let $G_j$ be the set of gates
of $F$ where two paths from $P_j$ intersect. Then $|G_j|\leq s_j$. 

Let $\rho$ be an assignment of $\ket{0}$ or $\ket{1}$ to the input variable
wires not in $\ov{S_j}$. We denote the resulting formula by
$F_\rho$. Thus $F_\rho$ computes a Boolean function
$f_\rho\colon\{0,1\}^{n_j}\longrightarrow\{0,1\}$ which is a subfunction
of $f$ and a member of $\Sigma_j$. Consider a path 
\begin{equation}
\pi=(g_1,g_2,\ldots,g_m), \qquad m>2, \label{path} 
\end{equation} 
in $F_\rho$, where $g_1$ is an input wire or a gate in $G_j$, $g_m$ is a
gate in $G_j$ or the output wire of $F$, and $g_\ell\not\in G_j$ for
$1<\ell<m$. 

\begin{figure} \begin{center} \unitlength=.4mm 

\begin{picture}(150,20)(0,-20)
\put(0,0){\vector(1,0){30}}\put(0,16){\vector(1,0){30}}
\put(40,8){\makebox(0,0){\framebox(20,24){$\gamma_\ell$}}}
\put(10,5){\makebox(0,0){$q_2$}}\put(10,21){\makebox(0,0){$q_1$}}
\put(50,16){\vector(1,0){30}}
\put(90,8){\makebox(0,0){\framebox(20,24){$\gamma_{\ell+1}$}}}
\put(50,0){\vector(1,-1){30}}\put(80,0){\line(-1,-1){30}}
\put(100,16){\vector(1,0){30}}
\put(140,8){\makebox(0,0){\framebox(20,24){$\gamma_{\ell+2}$}}}
\put(100,0){\vector(1,-1){30}}\put(130,0){\line(-1,-1){30}}
\put(80,0){\vector(1,1){0}}\put(130,0){\vector(1,1){0}}
\put(102,-19){\makebox(0,0){$q_3$}}

\end{picture} 
\end{center} \caption{The qubits $q_1$ and $q_2$ are strong companions at
step $\ell$, the qubits $q_1$ and $q_3$ are companions at step $\ell+2$.}
\label{fig5} \end{figure}
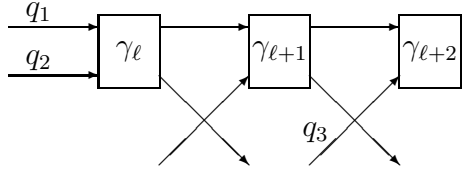 

To show how we can squeeze paths like (\ref{path}) (this is the essence
of the Nechiporuk's method), we introduce the following notations. We
consider a natural ordering $\gamma_1,\gamma_2,\ldots,\gamma_t$ on the
gates of the formula $F_\rho$, and regard $F_\rho$ as a computation in
$t$ steps where at step $\ell$ the corresponding gate $\gamma_\ell$ is
performed. We say two qubits $q_1$ and $q_2$ are {\em strong companions}
of each other at step $\ell$ if there is a gate $\gamma_j$ such that
$j\leq \ell$ and $q_1$ and $q_2$ are inputs of $\gamma_j$. We say qubits
$q_1$ and $q_2$ are {\em companions} of each other at step $\ell$ if
there exists a sequence $r_1,r_2,\ldots,r_p$ of qubits such that
$r_1=q_1$, $r_p=q_2$, and $r_j$ and $r_{j+1}$ (for $1\leq j\leq p-1$)
are strong companions of each other at step $\ell$ (see Figure \ref{fig5}). 
If $q_1$ and $q_2$
are companions at step $\ell$ then they are also companions at any step
after $\ell$. For a gate $g$, we define the {\em set of companions} of 
$g$ as the union of all companions of input qubits of $g$.

\begin{figure} \begin{center} \unitlength=.4mm 

\begin{picture}(260,90)(-60,-5)
\put(-60,32){\vector(1,0){30}}\put(-60,48){\vector(1,0){30}} 
\put(-50,38){\makebox(0,0){$q_1$}}\put(-50,54){\makebox(0,0){$q_0$}}
\put(20,48){\makebox(0,0){$q_0$}}
\put(-10,40){\makebox(0,0){\framebox(40,24){$g_1=\gamma_{j_0}$}}}
\put(10,42){\vector(1,0){30}} 
\put(10,32){\vector(1,-1){30}}
\put(50,40){\makebox(0,0){\framebox(20,24){$g_2$}}} 
\put(10,42){\vector(1,0){30}} 
\put(75,40){\circle*{3}}\put(82,40){\circle*{3}}\put(89,40){\circle*{3}}
\put(116,40){\makebox(0,0){\framebox(24,24){$\scriptsize g_{m-1}$}}}
\put(178,40){\makebox(0,0){\framebox(40,24){$g_m=\gamma_{j_1}$}}}
\put(128,42){\vector(1,0){30}}\put(143,47){\makebox(0,0){$q_0$}} 
\put(82,32){\vector(1,-1){30}}\put(128,32){\vector(1,-1){30}}
\put(50,15){\circle*{3}}\put(57,15){\circle*{3}}\put(64,15){\circle*{3}}
\put(113,15){\circle*{3}}\put(120,15){\circle*{3}}\put(127,15){\circle*{3}}
\put(40,47){\line(-1,1){30}}\put(158,47){\line(-1,1){30}}
\put(82,48){\line(-1,1){30}}\put(40,47){\vector(1,-1){0}}
\put(158,47){\vector(1,-1){0}}\put(82,48){\vector(1,-1){0}}
\put(153,65){\makebox(0,0){$q_2$}}
\put(95,65){\circle*{3}}\put(102,65){\circle*{3}}\put(109,65){\circle*{3}}
\put(37,65){\circle*{3}}\put(44,65){\circle*{3}}\put(51,65){\circle*{3}}
\put(100,-12){\makebox(0,0){${\cal M}_\pi$}}
\put(62,92){\makebox(0,0){$Q_\pi$}}
\put(62,81){\makebox(0,0){$\makebox[1.8in]{\downbracefill}$}}
\put(100,-3){\makebox(0,0){$\makebox[2in]{\upbracefill}$}}

\end{picture} 
\end{center} \caption{Squeezing a path.}\label{fig6} \end{figure}
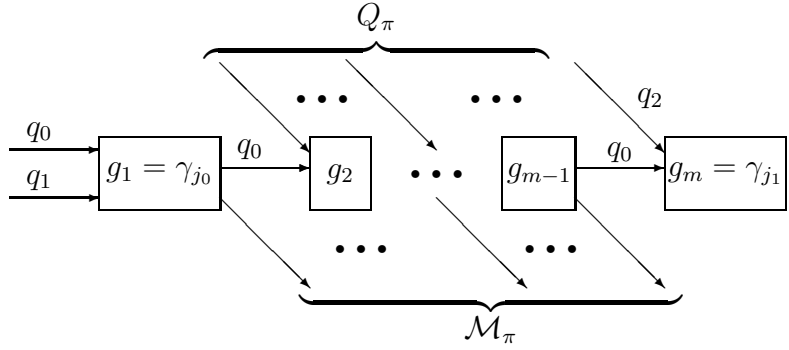

Suppose that in the path (\ref{path}) $g_1=\gamma_{j_0}$,
$g_m=\gamma_{j_1}$, the inputs of $g_1$ are $q_0$ and $q_1$,
the output of $\gamma_{j_0}$ from the path
(\ref{path}) is the qubit $q_0$, and the input of $\gamma_{j_1}$ not from
the path (\ref{path}) is the qubit $q_2$. Note that $q_0$ is the
companion of $q_2$ at step $j_1$. Let $Q_\pi$ be union of all sets of
companions of $g_1,\ldots,g_{m-1}$ at step $j_1$ minus $q_0$ and $q_1$
(see Figure \ref{fig6}). 
Let $C_0$ be the circuit defined by the gates $g_1,\ldots,g_{m-1}$ from the 
path (\ref{path}). Suppose that $|Q_\pi|=v$ and consider $C_0$ as an operation
acting on ${\cal H}=\cc^2\otimes\cc^2\otimes\cc^{2^v}$, where
$\ket{\alpha_0}\otimes\ket{\alpha_1}\otimes\ket{\alpha}\in{\cal H}$
denotes the state of $q_0$, $q_1$, and the companion qubits in $Q_\pi$,
respectively. Note that all qubits in $Q_\pi$ are constant inputs of
$F_\rho$ and do not intersect any other path like (\ref{path}), because
$F$ is a formula. So the input $\ket{\alpha}$ of the subcircuit $C_0$ is 
the same for all $\ket{\alpha_0}$ and $\ket{\alpha_1}$. Therefore, we could 
substitute the subcircuit $C_0$ by $\widetilde{C_0}$ such that on input
$\ket{\alpha_0}\ket{\alpha_1}\ket{0\cdots0}$, the subcircuit $\widetilde{C_0}$
first computes $\ket{\alpha_0}\ket{\alpha_1}\ket{\alpha}$ then applies the action
of $C_0$. Suppose that the act of $\widetilde{C_0}$ be defined as follows 
\begin{equation}
\ket{\alpha_0}\otimes\ket{\alpha_1}\otimes\ket{0\cdots 0}\longrightarrow
\sum_{c_0,c_1\in\{0,1\}}\ket{c_0}\otimes\ket{c_1}\otimes
\ket{A_{c_0,c_1}^{\alpha_0,\alpha_1}} , \label{unitary1} 
\end{equation}
where $\alpha_0,\alpha_1\in\{0,1\}$ and
$\ket{A_{c_0,c_1}^{\alpha_0,\alpha_1}}\in\cc^{2^v}$ may be not a unit
vector. Let ${\cal A}_\pi\subseteq \cc^{2^v}$ be the vector space spanned by
$\ket{A_{c_0,c_1}^{\alpha_0,\alpha_1}}$, 
for $\alpha_0,\alpha_1,c_0,c_1\in\{0,1\}$ and $d=\dim({\cal A}_\pi)$. Then
$1\leq d\leq 16$. Let $\ket{A_1^\pi},\ldots,\ket{A_d^\pi}$ be an orthonormal
basis for ${\cal A}_\pi$. Then we can rewrite (\ref{unitary1}) as follows
\begin{equation} 
\ket{\alpha_0}\otimes\ket{\alpha_1}\otimes\ket{0\cdots
0}\longrightarrow \sum_{c_0,c_1\in\{0,1\}}\sum_{1\leq j\leq
d}\lambda_{j,c_0,c_1}^{\alpha_0,\alpha_1}
\ket{c_0}\otimes\ket{c_1}\otimes\ket{A_j^\pi} . \label{unitary2}
\end{equation}

Let ${\cal M}_\pi$ be the set of those unitary operations that are 
performed after one of the gates $g_1,\ldots,g_{m-1}$ on some qubits in $Q_\pi$
before the step $j_1$. Since qubits in $Q_\pi$ do not interact with any 
other path of the form (\ref{path}), by Lemma \ref{composition} (b), we can postpone
all operations in ${\cal M}_\pi$ after we computed the output
of $g_m$. Let $\pi_1,\ldots,\pi_k$ be a natural ordering on the paths in $P_j$
(i.e., the last gate of $\pi_{j+1}$ is not performed before the last gate of $\pi_j$).
Consider the sets of postponed operations ${\cal M}_{\pi_1},\ldots,{\cal M}_{\pi_k}$.
Once again Lemma \ref{composition} implies that we can postpone operations in
${\cal M}_{\pi_1}$ after the last gate of $\pi_2$, and so on. Repeating this argument
shows that we can postpone all operations in ${\cal M}_{\pi_1},\ldots,{\cal M}_{\pi_k}$
after we compute the output qubit. In this way, the state of the output qubit, before
the postponed operations ${\cal M}_{\pi_1},\ldots,{\cal M}_{\pi_k}$ is applied,
is of the form
\begin{equation} 
\ket{0}\otimes\ket{M}+\ket{1}\otimes\ket{N} ,
\label{unitary3} \end{equation} 
where the first qubit is the output
qubit and $\ket{M}$ and $\ket{N}$ are superpositions of tensor
products of orthonormal vectors $\ket{A_k^{\pi_j}}$ used in (\ref{unitary2}). By
Lemma \ref{basis}, these tensor products of the vectors $\ket{A_k^{\pi_j}}$ are
unit vectors and pairwise orthogonal. The unitary operations in the sets
${\cal M}_{\pi_j}$ (for paths $\pi_j$ of the form (\ref{path})), which are postponed
to the end, do not change the lengths of $\ket{M}$ and $\ket{N}$. Thus,
as far as the computation of the Boolean function $f_\rho$ is concerned,
we can ignore all the postponed unitary operations. For this reason we
construct the circuit $\ov{F_\rho}$ from the formula $F_\rho$ by
eliminating all postponed operations in ${\cal M}_{\pi_j}$, substituting for 
each path $\pi_j$ of the form (\ref{path}) the companion qubits in $Q_{\pi_j}$ 
by four new qubits, and the
unitary operation (\ref{unitary2}) by the operation defined as
\begin{equation}
\ket{\alpha_0}\otimes\ket{\alpha_1}\otimes\ket{0000}\longrightarrow
\sum_{c_0,c_1\in\{0,1\}}\sum_{0\leq j\leq
15}\lambda_{j,c_0,c_1}^{\alpha_0,\alpha_1}
\ket{c_0}\otimes\ket{c_1}\otimes\ket{j} . \label{unitary4}
\end{equation} The output of the circuit $\ov{F_\rho}$, instead of
(\ref{unitary3}), is of the form \begin{equation}
\ket{0}\otimes\ket{M'}+\ket{1}\otimes\ket{N'} , \label{unitary5}
\end{equation} where $\norm{\ket{M}}=\norm{\ket{M'}}$ and
$\norm{\ket{N}}=\norm{\ket{N'}}$. So the circuit $\ov{F_\rho}$ computes
$f_\rho$. Moreover, \[ \mbox{\sf size}(\ov{F_\rho}) =O(s_j), \] and for
another assignment $\tau$, the corresponding circuit $\ov{F_{\tau}}$
differs from $\ov{F_\rho}$ only at unitary operations defined by
(\ref{unitary4}). 

The above discussion implies that $\sigma_j$, the number of subfunctions
on $S_j$, is at most the number of different Boolean functions computed
by size $O(s_j)$ quantum circuits. Therefore, by Theorem \ref{knill}, we
get \[ \sigma_j\leq 2^{O(s_j\log s_j)} .\] So
$s_j=\Omega({\log(\sigma_j)}/{\log\log(\sigma_j)})$. Now the theorem
follows from (\ref{size}). $\Box$ 

\vspace{6mm} To apply the general bound of the above theorem, we could
consider any of the several explicit functions used in the case of
Boolean formulas (see \cite{dunne} and \cite{wegener}). As we mentioned in 
the beginning of this section, we consider the Element Distinctness function
$\mbox{ED}_n$. For this function
$\sigma_j>\ell^{\ell-1}$, where $\ell=\Omega(n/\log n)$. Therefore, we get 
the lower bound $\Omega(\ell^2)=\Omega(n^2/\log^2 n)$ for the formula size. 

\begin{th} Any quantum formula computing $\mbox{\em ED}_n$ has size
$\Omega(n^2/\log^2 n)$. \end{th} 

\section{Concluding Remarks} 

We extend a classical technique for proving lower bound for Boolean
formula size to quantum formulas. The difficult part was to effectively deal
with the phenomenon of entanglement of qubits. While we have been successful
in extending a classical technique to the quantum case, the challenges 
encountered indicate that in general the problem of extending methods of Boolean
case to the quantum case may not have simple solutions. 
For example, even the seemingly simple issue of the exact 
relationship between quantum formulas and quantum circuits has not been resolved.
In the Boolean
case, simulation of circuits by formulas is a simple fact, but in the
quantum case it is not clear whether every quantum circuit can be simulated by a quantum
formula. In particular, it is not clear that in the process of going from quantum
circuits to formulas, how we can modify the underlying entanglement of
qubits while keeping the probability of reaching to the final
answer the same.

\appendix
\section{Appendix: Counting the number of Boolean functions computed by quantum circuits
of a given size}

In this appendix we prove the following upper bound.

\begin{th} 
The number of different $n$--variable Boolean functions that 
can be computed by size $N$ quantum circuits ($n\leq N$) with $d$--input 
$d$--output elementary gates (for some constant $d$)
is at most $2^{O(nN)+O(N\log N)}$.  
\end{th} 

Our proof is based on Warren's bound on the number of different sign--assignments to
real polynomials \cite{warren}. We begin with some necessary notations.

Let $P_1(x_1,\ldots,x_t),\ldots,P_m(x_1,\ldots,x_t)$ be real polynomials. 
A {\em sign--assignment} to these polynomials is a system of inequalities
\begin{equation}
  P_1(x_1,\ldots,x_t)\,\Delta_1\, 0,\ldots,P_m(x_1,\ldots,x_t)\,\Delta_m\, 0 , 
\label{sign}
\end{equation}
where each $\Delta_j$ is either ``$<$'' or ``$>$''. The sign--assignment (\ref{sign}) is
called {\em consistent} if this system has a solution in $\rr^t$.

\begin{th} {\em (Warren \cite{warren})}
Let $P_1(x_1,\ldots,x_t),\ldots,P_m(x_1,\ldots,x_t)$ be real polynomials, each of degree at 
most $d$. Then there are at most $(4edm/t)^t$ consistent sign--assignments of the form
{\em (\ref{sign})}.
\label{warrentheo}
\end{th}

We are now ready to prove Theorem \ref{knill}. We consider the class of quantum circuits 
of size $N$ with $d$--bit gates computing $n$--variable Boolean functions. 
Without loss of generality, we can assume that $n'$, the number of input wires 
of such circuits, is at most $d\cdot N$. We define an equivalence relation
$\Join$ on such circuits: we write $C_1\Join C_2$ if and only if
$C_1$ and $C_2$ differ only in
the label of their gates; in another word, $C_1$ and $C_2$ have the same underlying
graph but the corresponding gates in these circuits may compute different unitary
operations. The number of different equivalence classes is at most
\[ \binom{n'}{d}^{N} \leq (dN)^{dN} = 2^{O(N\log N)}.\]
Now we find an upper bound for the number of different Boolean functions that can be computed
by circuits in the same equivalence class. Fix an equivalence class $\cal E$.
We use the variables
$a_1+ib_1,a_2+ib_2,\ldots,a_\mu+ib_\mu$, where $\mu=2^{2d}N$, to denote the 
entries of the matrices of
the gates of a circuit $C$ in $\cal E$. By substituting appropriate values to the
variables $a_1,\ldots,a_\mu,b_1,\ldots,b_\mu$, we get all circuits in $\cal E$.
On input $\alpha=(\alpha_1,\ldots,\alpha_n)\in\{0,1\}^n$, the probability
that $C$ outputs 1 can be represented by a {\em real} polynomial 
$P_\alpha(a_1,\ldots,a_\mu,b_1,\ldots,b_\mu)$. The degree of $P_\alpha$ is at most $N^2$.
There are $2^n$ polynomials $P_\alpha$ and the number of different Boolean functions
can be computed by $C$ by changing the unitary operators of its gates is at most the
number of different consistent sign--assignments to the following system:
\[ \mbox{$P_\alpha(a_1,\ldots,a_\mu,b_1,\ldots,b_\mu)-\frac{2}{3}$ and
$P_\alpha(a_1,\ldots,a_\mu,b_1,\ldots,b_\mu)-\frac{1}{3}$},\qquad \alpha\in\{0,1\}^n .\]
By Theorem \ref{warrentheo} this number is bounded from the above by
\[\biggl ( \frac{4eN^22^{n+1}}{2\mu} \biggr )^{2\mu}=2^{O(nN)+O(N\log N)} .\quad\Box  \]


\begin{thebibliography}{11} 

\iffalse
\bibitem{adelman} L. M. Adelman, J. Demarrias, and M. A. Huang,
``Quantum computability,'' %{\em SIAM J. Computing}, 26(1997), pp.
1524--1540. 

\bibitem{barenco} A. Barenco, ``A universal two--bit gate for quantum
computation,'' %{\em Proc. Roy. Soc. London Ser. A.}, 449(1995), pp.
679--683. 
\fi

\bibitem{barenco-etal} A. Barenco, C. Bennett, R. Cleve, D.
DiVincenzo, N. Margolus, P. Shor, T. Sleator, J. Smolin, and H.
Weinfurter, ``Elementary gates for quantum computation,'' {\em Phys.
Rev. A}, 52(1995), pp. 3457--3467. 

\bibitem{beals}
 R. Beals, H. Buhrman, R. Cleve, M. Mosca, R. de Wolf,
Quantum lower bounds by polynomials, in {\em Proceedings 39th IEEE Annual 
Symposium on Foundations of Computer Science},  pp. 352-361, 1998.

\bibitem{vazirani} E. Bernstein and U. Vazirani, ``Quantum complexity
theory,'' {\em SIAM J. Computing}, 26(1997), pp. 1411--1473. 

\bibitem{boppana} R. B. Boppana and M. Sipser, ``The complexity of
finite functions,'' in {\em Handbook of Theoretical Computer Science},
Vol. A, ``Algorithms and Complexity'' (J. van Leeuwen, Ed.), pp.
757--804, Elsevier Science, New York, MIT Press, Cambridge, MA, 1990. 

\bibitem{deutsch85} D. Deutsch, ``Quantum theory, the Church--Turing
principle and the universal quantum computer,'' {\em Proc. Roy. Soc.
London Ser. A.}, 400(1985), pp. 97--117. 

\bibitem{deutsch89} D. Deutsch, ``Quantum computational networks,'' {\em
Proc. Roy. Soc. London Ser. A.}, 425(1989), pp. 73--90. 

\iffalse
%\bibitem{deutsch95} %D. Deutsch, A Barenco and A. Ekert, ``
Universality in quantum computation,'' %{\em Proc. Roy. Soc. London Ser.
A.}, 449(1995), pp. 669--677. 

%\bibitem{deutschjozsa} %D. Deutsch and R. Jozsa, ``Rapid solution of
problems by quantum computation,'' %{\em Proc. Roy. Soc. London Ser.
A.}, 439(1992), pp. 553--558. 

%\bibitem{divincenzo} %D. P. DiVincenzo, ``Two--bit gates are universal
for quantum computation,'' %{\em Phys. Rev. A}, 51(1995), pp.
1015--1022. 
\fi

\bibitem{dunne} P. E. Dunne, {\em The Complexity of Boolean Networks},
Academic Press, London, 1988. 
\iffalse
%\bibitem{feynman} %R. Feynman, ``Simulating physics with computers,''
{\em Internat. J. Theoret. %Phys.}, 21(1982), pp. 467--488. 
\fi
\bibitem{grover} L. Grover, ``A fast quantum mechanical algorithm for
database search,'' in {\em Proceedings of 28th ACM Symposium on Theory
of Computing}, pp. 212--219, 1996. 

\bibitem{knill} E. Knill, ``Approximating by quantum circuits,'' LANL
e--print quant--ph/9508006. 
\iffalse
%\bibitem{lloyd} %S. Lloyd, ``Almost any quantum logic gate is
universal,'' %{\em Phys. Rev. Lett.}, 75(1995), pp. 346--349. 

%\bibitem{reck} %M. Reck, A. Zeilinger, H. J. Bernstein and P. Bertani,
``Experimental %realization of any discrete unitary operator,'' {\em
Phys. Rev. Lett.} %73(1995), pp. 58--61. 
\fi
\bibitem{shor} P. Shor, ``Polynomial--time algorithms for prime
factorization and discrete logarithms on a quantum computer,'' {\em SIAM
J. Computing}, 26(1997), pp. 1484--1509. 

\bibitem{simon} D. Simon, ``On the power of quantum computation,'' {\em
SIAM J. Computing}, 26(1997), pp. 1474--1483. 

\iffalse
%\bibitem{sleator} %T. Sleator and H. Weinfurter, ``Realizable universal
quantum logic gates,'' %{\em Phys. Rev. Lett.}, 74(1995), 4087--90. 
\fi

\bibitem{turan} Gy. Tur\'{a}n and F. Vatan, ``On the computation of
Boolean functions by analog circuits of bounded fan--in,'' {\em J.
Computer and System Sciences,} 54(1997), pp. 199--212. 

\bibitem{warren}
H. E. Warren, ``Lower bounds for approximating by non--linear manifolds,'' 
{\em Trans. Amer. Math. Soc.}, 133(1968), pp. 167--178.

\bibitem{wegener} I. Wegener, {\em The Complexity of Boolean Functions},
Teubner--Wiley, New York, 1987. 

\bibitem{yao} A. Yao, ``Quantum circuit complexity,'' in {\em Proc. 34th
IEEE Symposium on Foundations of Computer Science}, 1993, pp. 352--361. 

\end{thebibliography}
\end{document}